\documentclass{article}
\usepackage{spconf,amsmath,graphicx}

\usepackage{hyperref}       % hyperlinks
\usepackage{url}            % simple URL typesetting
\usepackage{enumitem}
\usepackage{booktabs}
\usepackage{graphicx}
\usepackage{booktabs}
\usepackage{threeparttable}
\usepackage{subcaption}
\usepackage{tabularx}
\usepackage{todonotes}
\usepackage{lipsum}  % for dummy text
\setlist{nosep, leftmargin=14pt}
\usepackage{stfloats}
\usepackage{mwe} % to get dummy images
\usepackage{float}
\usepackage{multicol}
\title{A Comprehensive Dataset and Automated Pipeline for Nailfold Capillary Analysis}
\name{\begin{tabular}{c}
Linxi Zhao$^{1}$, Jiankai Tang$^{1}$, Dongyu Chen$^{1}$, Xiaohong Liu$^{2}$, Yong Zhou$^{3}$, \\
Yuanchun Shi$^{1}$, Guangyu Wang$^{4}$, Yuntao Wang$^{\star}$$^{1}$
\end{tabular}\thanks{$^{\star}$Corresponding author}}

\address{
$^{1}$Key Laboratory of Pervasive Computing, Ministry of Education, \\ Department of Computer Science and Technology, Tsinghua University, Beijing, China\\
$^{2}$UCL Cancer Institute, University College London, London, UK\\
$^{3}$Clinical Research Institute, Shanghai General Hospital, Shanghai, China\\
$^{4}$State Key Laboratory of Networking and Switching Technology, \\
Beijing University of Posts and Telecommunications, Beijing, China
}

\begin{document}
%\ninept
%
\maketitle
\begin{abstract}
Nailfold capillaroscopy is widely used in assessing health conditions, highlighting the pressing need for an automated nailfold capillary analysis system. In this study, we present a pioneering effort in constructing a comprehensive nailfold capillary dataset—321 images, 219 videos from 68 subjects, with clinic reports and expert annotations—that serves as a crucial resource for training deep-learning models. Leveraging this dataset, we finetuned three deep learning models with expert annotations as supervised labels and integrated them into a novel end-to-end nailfold capillary analysis pipeline. This pipeline excels in automatically detecting and measuring a wide range of size factors, morphological features, and dynamic aspects of nailfold capillaries. 
We compared our outcomes with clinical reports. Experiment results showed that our automated pipeline achieves an average of sub-pixel level precision in measurements and $89.9\%$ accuracy in identifying morphological abnormalities.  
These results underscore its potential for advancing quantitative medical research and enabling pervasive computing in healthcare.
Our data and code are available at {\small\texttt{\url{https://github.com/THU-CS-PI-LAB/ANFC-Automated-Nailfold-Capillary}}}.

\end{abstract}
\begin{keywords}
Nailfold capillaroscopy dataset, automated analysis pipeline
\end{keywords}
\vspace{-5px}
\section{Introduction}
\vspace{-5px}
\label{sec:intro}
Nailfold capillaroscopy, a traditional imaging technique for health condition assessment, is valued for its non-invasive, cost-effective, and user-friendly features. 
It has shown great potential in the diagnosis of cardiovascular and immune diseases in the last 30 years\cite{hahn1998hemodynamics,etehad2015nailfold,SMITH2023101849,smith2020standardisation,arvanitaki2021peripheral,mishra2021nailfold}.
Previous research has identified statistically significant differences in quantitative features from capillaroscopy images between patients and those with conditions such as Systemic Scleroderma and rheumatism\cite{hahn1998hemodynamics,berks2014automated,bharathi2023deep}. However, manually applying this method poses challenges as it demands substantial human effort for the measurement of morphological features and relies on subjective, experience-based assessment\cite{karbalaie2019practical,el2022nailfold,tello2022challenge}.

Many existing clinical approaches rely on traditional image-processing techniques, offering superior speed compared to conventional manual assessments\cite{berks2014automated,bourquard_analysis_2015,berks_automated_2018}. The introduction of machine learning marks a pivotal shift, presenting automated medical image analysis as a promising alternative due to its higher accuracy compared to traditional image processing algorithms\cite{tello2022challenge}. Recent studies have built task-specific deep-learning models and algorithms such as nailfold capillary segmentation\cite{bharathi2023deep,hariyani2020capnet}, measurement of capillary size and density\cite{tello2022challenge}, and white cell counting\cite{kim_automated_2020}. 
Despite notable achievements, the untapped potential of fully automated medical image analysis remains for two key reasons: firstly, there is a critical shortage of large-scale, annotated datasets necessary for the effective training and fine-tuning of deep learning models; secondly, while previous studies have automated individual components, a fully automated system covering the entire process still awaits realization.

\vspace{-10px}
\begin{figure}[hbt]
\begin{minipage}[b]{1.0\linewidth}
  \centering
  \centerline{\includegraphics[width=8.5cm]{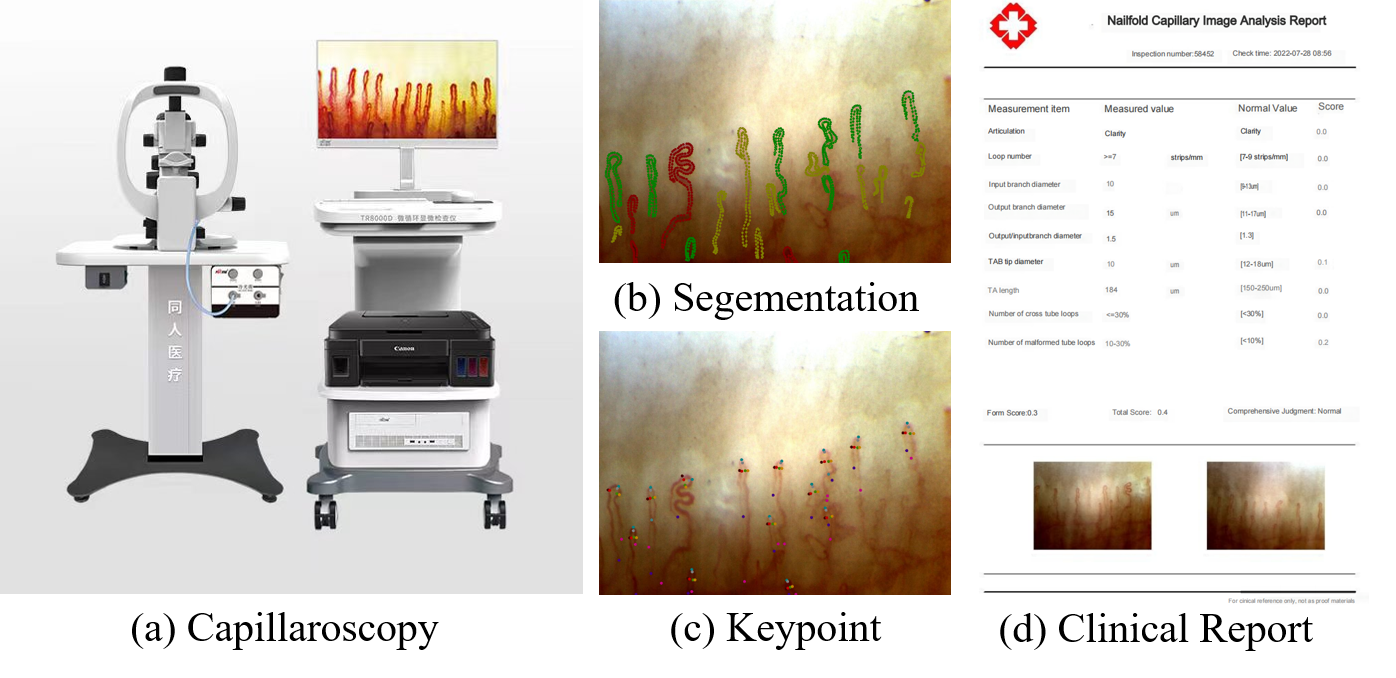}}
%  \vspace{2.0cm}
  \centerline{}\medskip
\end{minipage}
\vspace{-40px}

\caption{\textbf{Data Collection and Annotation.} (b) Segmentation with green, red, and yellow annotations for normal, enlarged, and blurred capillaries. (c) Keypoint annotation.}
% (a) Nailfold capillaroscopy. (b) Segmentation with green, red, and yellow for normal, enlarged, and blurred capillaries. (c) Keypoint annotation. (d) Clinical report.}
\label{fig:experiemnt}
\end{figure}

This paper addresses the gap by introducing a comprehensive dataset comprising 321 images and 219 videos of nailfold capillaries from 68 subjects, all annotated by experts under clinic guidance, along with 68 clinic reports. Leveraging our dataset, we build an innovative end-to-end nailfold capillary analysis pipeline, including preprocessing, segmentation, detection, and measurements. This automated system can detect and measure a wide array of morphological and dynamic features, covering almost all aspects applicable to medical diagnosis\cite{etehad2015nailfold,herrick2021quantitative}. Our automated nailfold capillary analysis system showcases high precision, making it well-suited for various quantitative medical research endeavors and the integration into pervasive computing systems designed to enhance human health\cite{cutolo2019nailfold}.
% --------------------------------------

\vspace{-10px}
\label{sec:pagestyle}

\begin{figure*}[hbt]
  \centering
  \begin{minipage}[b]{0.572\textwidth} %0.635
    \centering
    \includegraphics[width=\columnwidth, height=8.5cm, keepaspectratio]{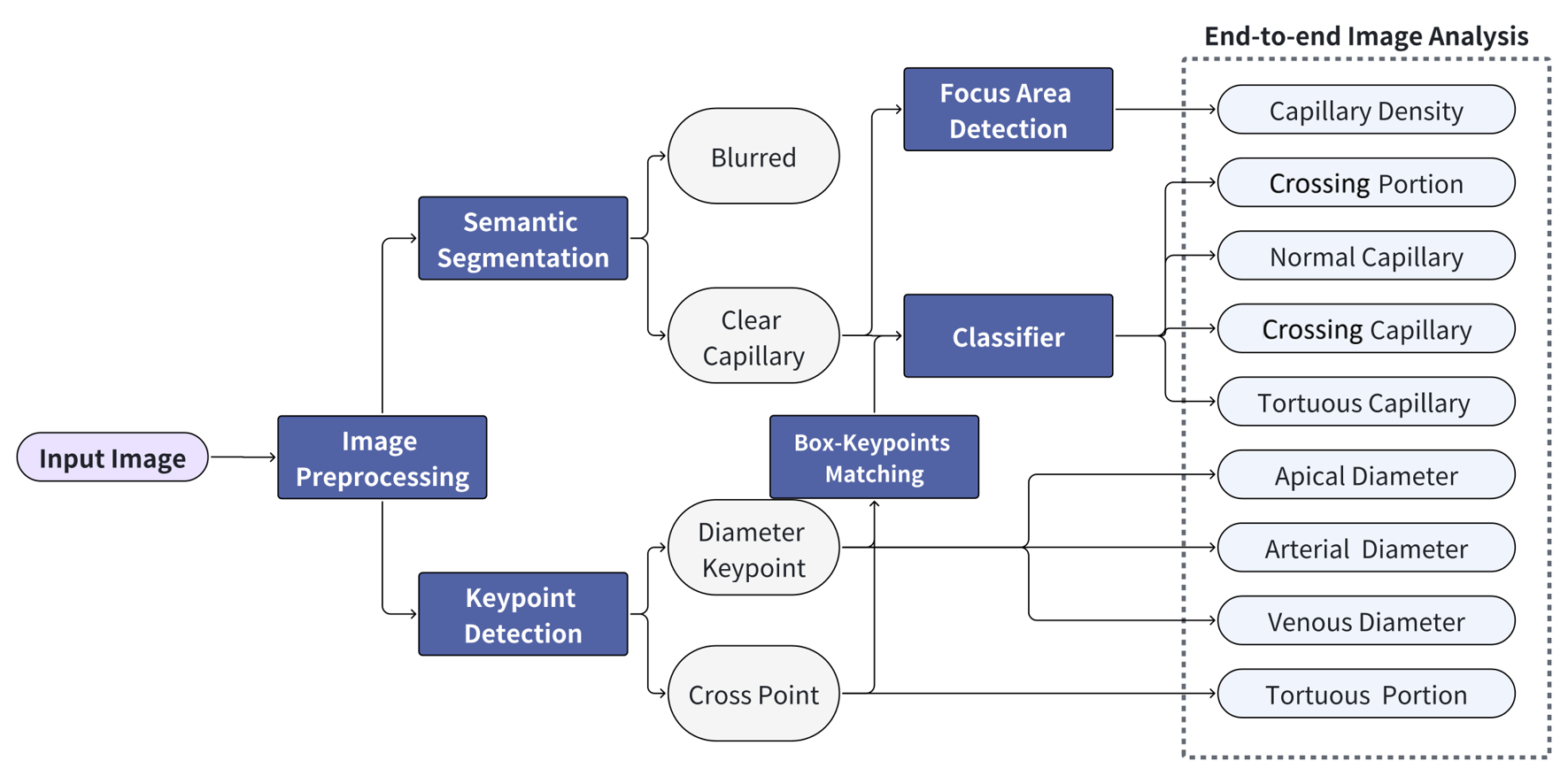}
    \subcaption{Nailfold Capillary Image End-to-end Analysis Pipeline}
  \end{minipage}%
  \hfill
  \begin{minipage}[b]{0.328\textwidth} %0.365
    \centering
    \includegraphics[width=\columnwidth, height=8.5cm, keepaspectratio]{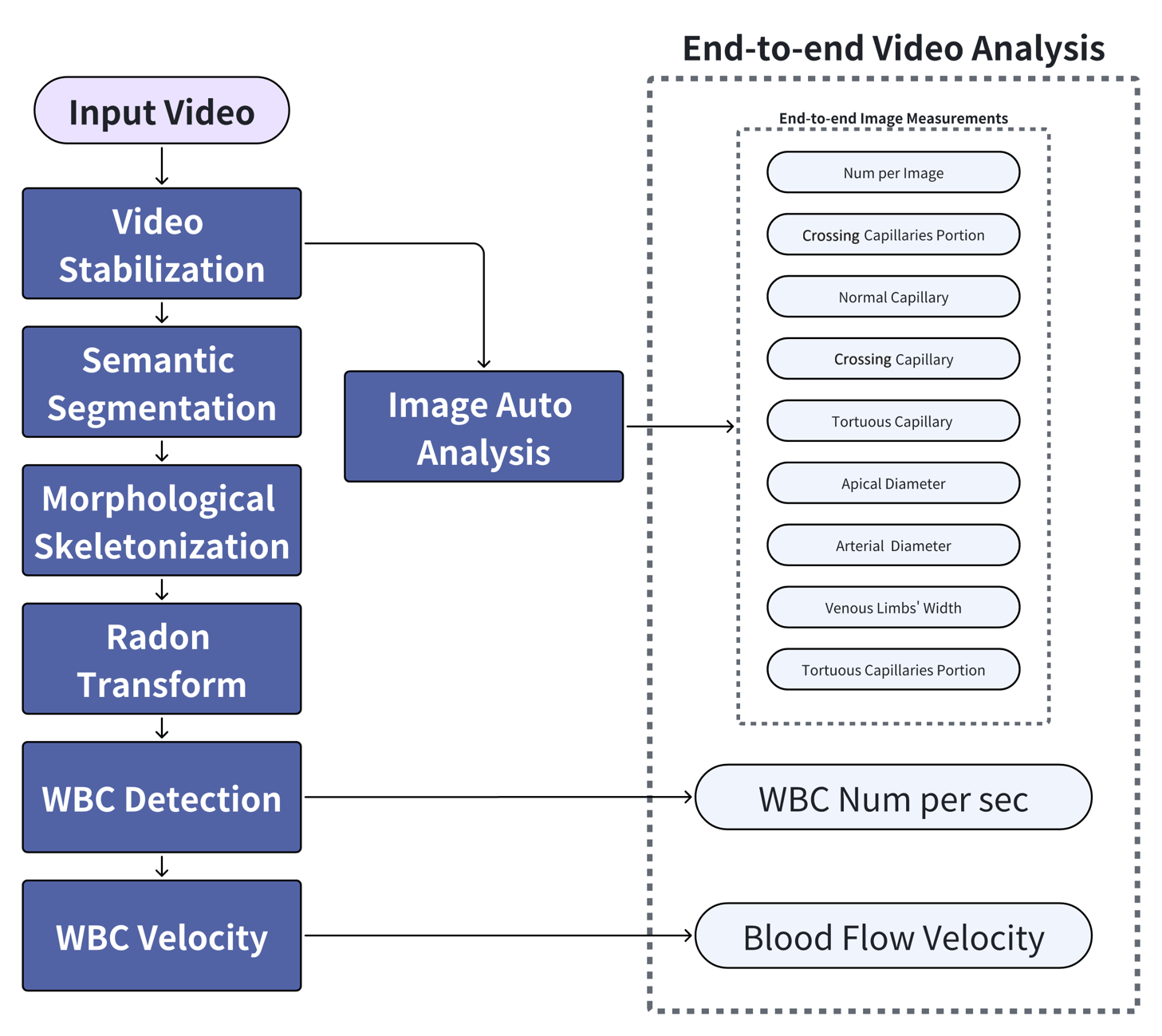}
    \subcaption{Nailfold Capillary Video End-to-end Analysis Pipeline}
  \end{minipage}
  \caption{Our algorithmic pipeline produces a comprehensive array of features vital for medical diagnosis, encompassing size factors, morphological features, and dynamic features.}
  \label{fig:pipeline}
\end{figure*}

 %----------
% \vspace{-5px}
\section{Dataset}
\vspace{-5px}
%subject 68, images 321， videos 219, frames 37321+321=37642.
%image shape 1024*768, video shape 1024*768 fps20.

Approved by the Institutional Review Board, we conducted the data collection experiment at Tangshan Hospital with TR8000D capillaroscopy, as Figure \ref{fig:experiemnt} illustrated. A total of 37,642 frames of nailfold capillaries from 68 subjects were captured, comprising 68 clinic reports, 321 images and 219 videos. Among the participants, there were 29 males and 36 females, with the age range spanning from 31 to 60 years. It is noteworthy that 3 participants opted not to disclose their gender and age information. All frames were shot at a resolution of $1,024 \times 768$, and the videos were recorded at a frame rate of 20 fps. Two samples of the nailfold capillary are shown in Figure \ref{fig:experiemnt} and a demo video is provided through this \href{https://github.com/THU-CS-PI-LAB/ANFC-Automated-Nailfold-Capillary}{link} for review.

\vspace{-10px}
\subsection{Data collection}
\vspace{-5px}
In the process of the experiment, participants are first asked to relax in the examination lab ($20-25^\circ$C) for at least five minutes. After hand washing and hand drying, a small amount of vegetable or baby oil is applied to one ring finger. The participant then places this finger under a nailfold capillaroscopy, as instructed by researchers, for the collection of microcirculation data. During the experiment, medical professionals would target the region of interest at the capillaroscopy and shoot 4-5 images and 3-4 videos for each subject.

\vspace{-10px}
\subsection{Data annotation}
\vspace{-5px}
The collected data underwent meticulous annotation and scoring, including medical report annotations and medical image feature markings, as Figure \ref{fig:experiemnt} illustrated. The medical reports, annotated manually by physicians, encompassed various parameters: clarity, capillary loop count (capillary density), afferent limb diameter, efferent limb diameter, apical diameter, hemorrhage, flow rate, white capillary microthrombi, exudation, capillary loop morphology, papilla, red blood cell aggregation, loop length, etc. Physicians provided additional scoring based on morphological, flow-related, and loop features. To train an automated nailfold analysis system, we employed professional annotators who, guided by medical guidelines, further marked all images. This included segmentation of the nailfold region and keypoint labeling of nailfold capillaries.

%----------
\begin{figure}[htb]

\begin{minipage}[h]{.49\linewidth}
  \centering
  \centerline{\includegraphics[width=3.5cm]{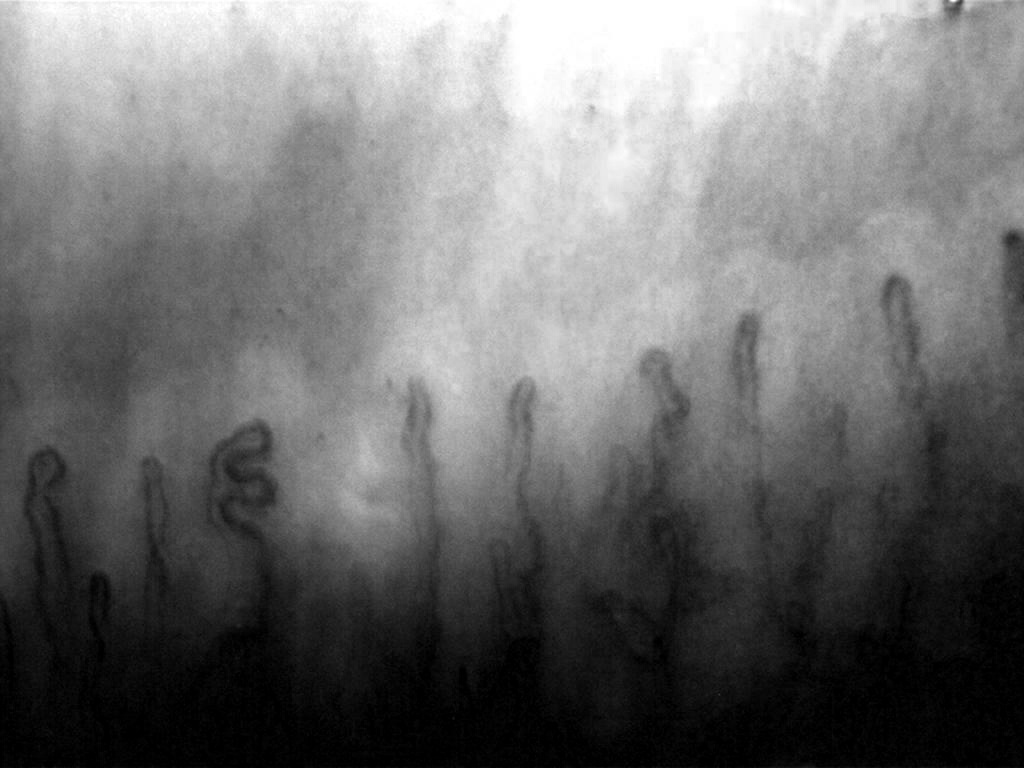}}
 % \vspace{1.5cm}
  \centerline{ (a) Preprocessed Image  }\medskip
\end{minipage}
% \hfill
\begin{minipage}[h]{0.49\linewidth}
  \centering
  \centerline{\includegraphics[width=3.5cm]{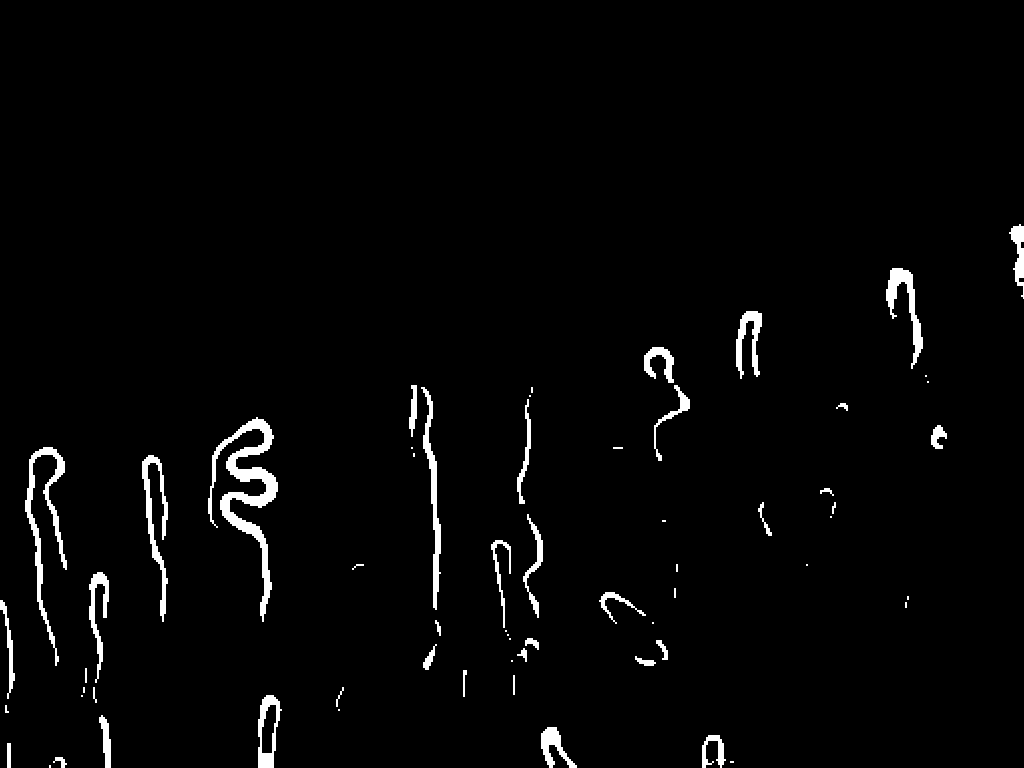}}
 % \vspace{1.5cm}
  \centerline{(b) Segmentation}\medskip
\end{minipage}
\begin{minipage}[h]{\linewidth}
  \centering
  \centerline{\includegraphics[width=8cm]{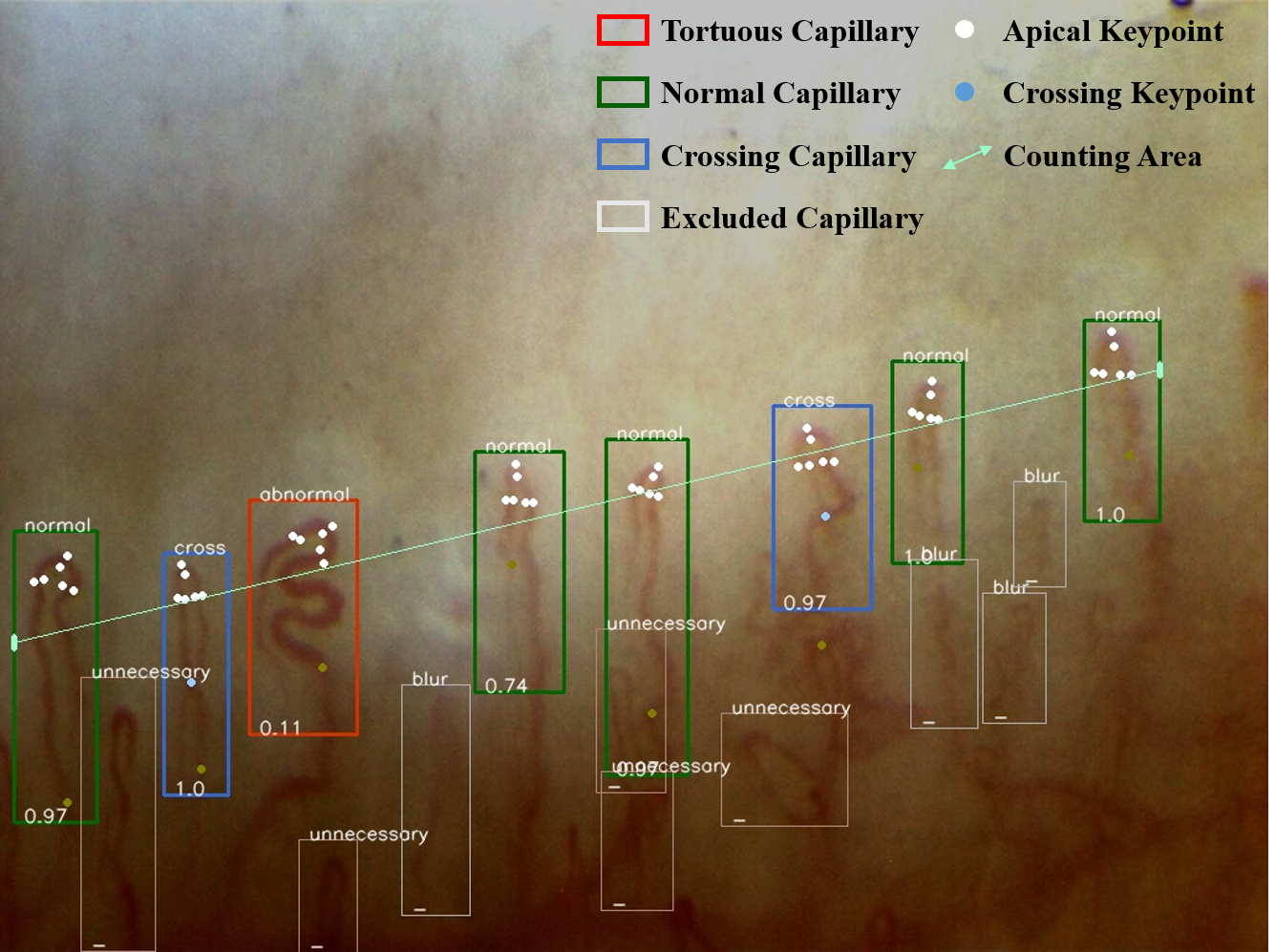}}
 % \vspace{2.0cm}
  \centerline{(c) Final Annotated Results}\medskip
\end{minipage}
 \vspace{-10px}
\caption{\textbf{Exemplifying the Stages of Our Nailfold Capillary Image Analysis System.} Excluded capillary refers to those initially proposed and subsequently excluded in the pipeline. }
\label{fig:case}
  \vspace{-15px}
\end{figure}
%----------
%----------

\section{Method and Results}

%---------
\vspace{-5px}
\subsection{End-to-end nailfold capillary image analysis pipeline}
\vspace{-5px}
The end-to-end nailfold capillary image analysis pipeline is illustrated in Figure \ref{fig:pipeline} and exemplified through a specific case presented in Figure \ref{fig:case}.

%----------
\begin{table*}[b]
    \centering
    \begin{threeparttable}
        \caption{Summary of Deep Learning Models, Datasets, and Evaluation Results}
        \small
        \begin{tabular}{ccccccc}
            \toprule
            \textbf{Model} & \textbf{Task} & \textbf{\#Dataset} & \textbf{Data Type} & \textbf{Train:Test Ratio}& \textbf{Eval Metric} & \textbf{Eval Score} \\
            \midrule
            U-Net\cite{ronneberger2015u} & Segmentation & 321 & Images & 4:1\tnote{1}& Sensitivity & 0.653 \\
            Mask R-CNN\cite{he2017mask} & Apical Keypoint Detection & 3,367 & Capillary Patch\tnote{2}& 7:3 & Sensitivity & 0.883 \\
            Mask R-CNN\cite{he2017mask} & Crossing Point Detection & 926 & Capillary Patch\tnote{2}& 7:3 & Sensitivity & 0.550 \\
            ResNet-18\cite{he2016deep} & Classification & 6,657\tnote{3}& Capillary Patch\tnote{2}& 3:1 & Accuracy & 0.800 \\
            \bottomrule
        \end{tabular}
        \begin{tablenotes}
            \item[1] The training set comprises 256 images across 54 subjects, while the test set includes 64 images from 14 distinct subjects, ensuring no overlap between the subjects in the training and test datasets. 
            \item[2] The image is cropped into patches, with each patch containing a single capillary.
            \item[3] The dataset includes 6,657 patches, categorized as normal capillary ($45.6\%$), tortuous capillary ($1.0\%$), blurred capillary ($39.5\%$), enlarged capillary ($6.4\%$), and hemorrhage ($7.5\%$).
        \end{tablenotes}
        \label{tab:model_metrics}
    \end{threeparttable}
\end{table*}

Initially, the input images are standardized and converted to grayscale during preprocessing. Subsequently, a fine-tuned U-Net model\cite{ronneberger2015u} is employed to extract nailfold semantic segmentation, enhancing the understanding of morphological features. Capillaries exhibiting clear structures are kept for further measurements, while those that are blurred, out-of-focus, or concealed are systematically filtered. Notably, images lacking sufficient clear capillaries are marked with NaN values in the final report. In parallel, a Mask R-CNN model\cite{he2017mask} identifies crossing points and detects keypoints for measurements which locates the apical and each capillary arterial and venous limb. A matching algorithm is then applied to ensure consistency between keypoints and detection boxes. In the final stage, all clear and matched capillaries within the counting area are classified by a pretrained ResNet-18\cite{he2016deep} into three categories—normal, crossing, and tortuous capillaries—aligned with relevant medical research. 

Our analysis system generates a comprehensive set of features crucial for medical diagnosis. This includes (1) size factors like the diameter of each capillary's apical, arterial, and venous limbs, (2) morphological features like the proportions of crossing and tortuous capillaries, and capillary density. 

\vspace{-10px}
\subsection{End-to-end nailfold capillary video analysis pipeline}
\begin{figure}[tbp]

% \hfill

\begin{minipage}[b]{0.49\linewidth}
  \centering
  \centerline{\includegraphics[width=\textwidth]{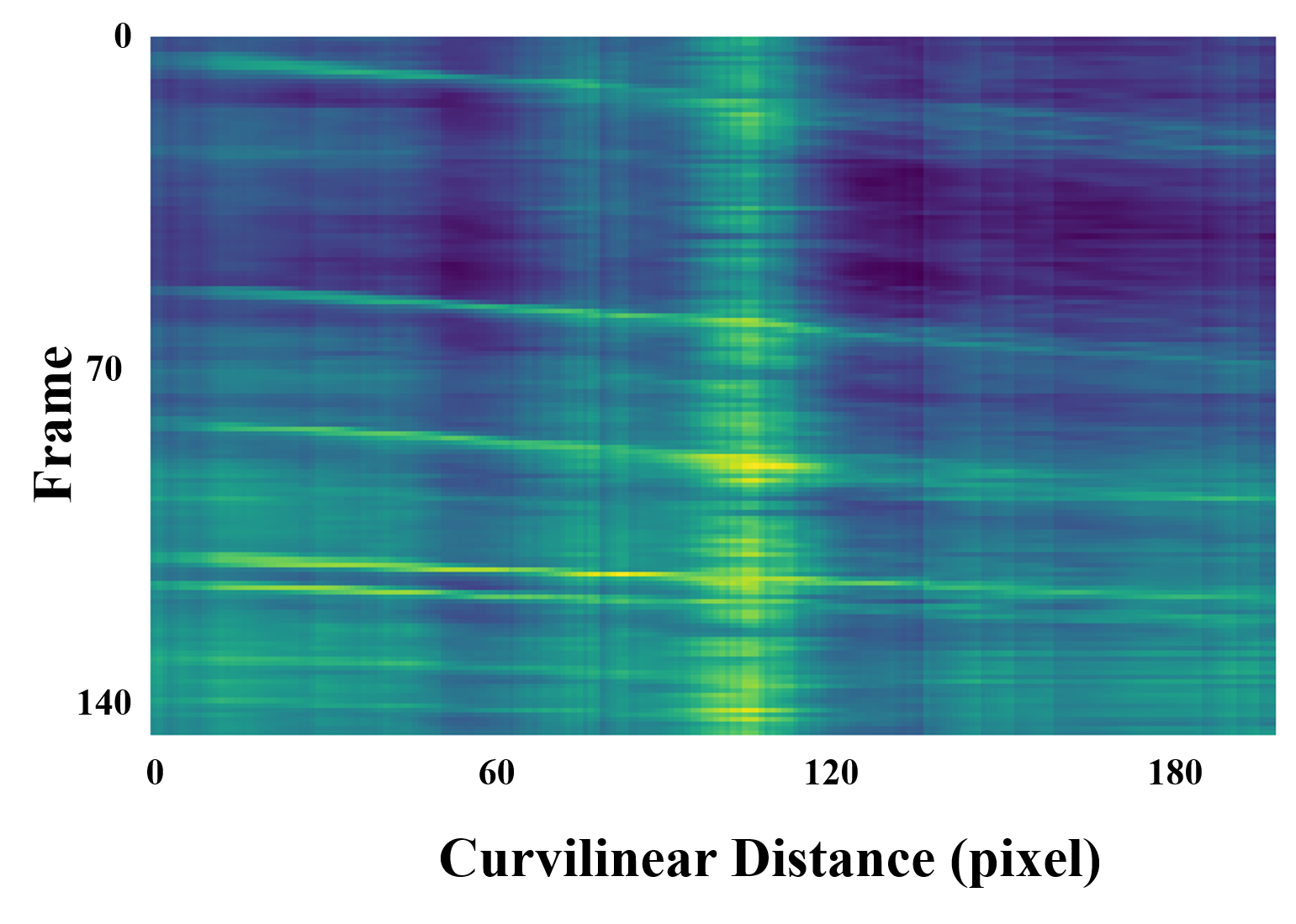}}
 % \vspace{1.5cm}
  \centerline{(a) Spatio-Temporal Profile}\medskip
\end{minipage}
\begin{minipage}[b]{0.49\linewidth}
  \centering
  \centerline{\includegraphics[width=\textwidth]{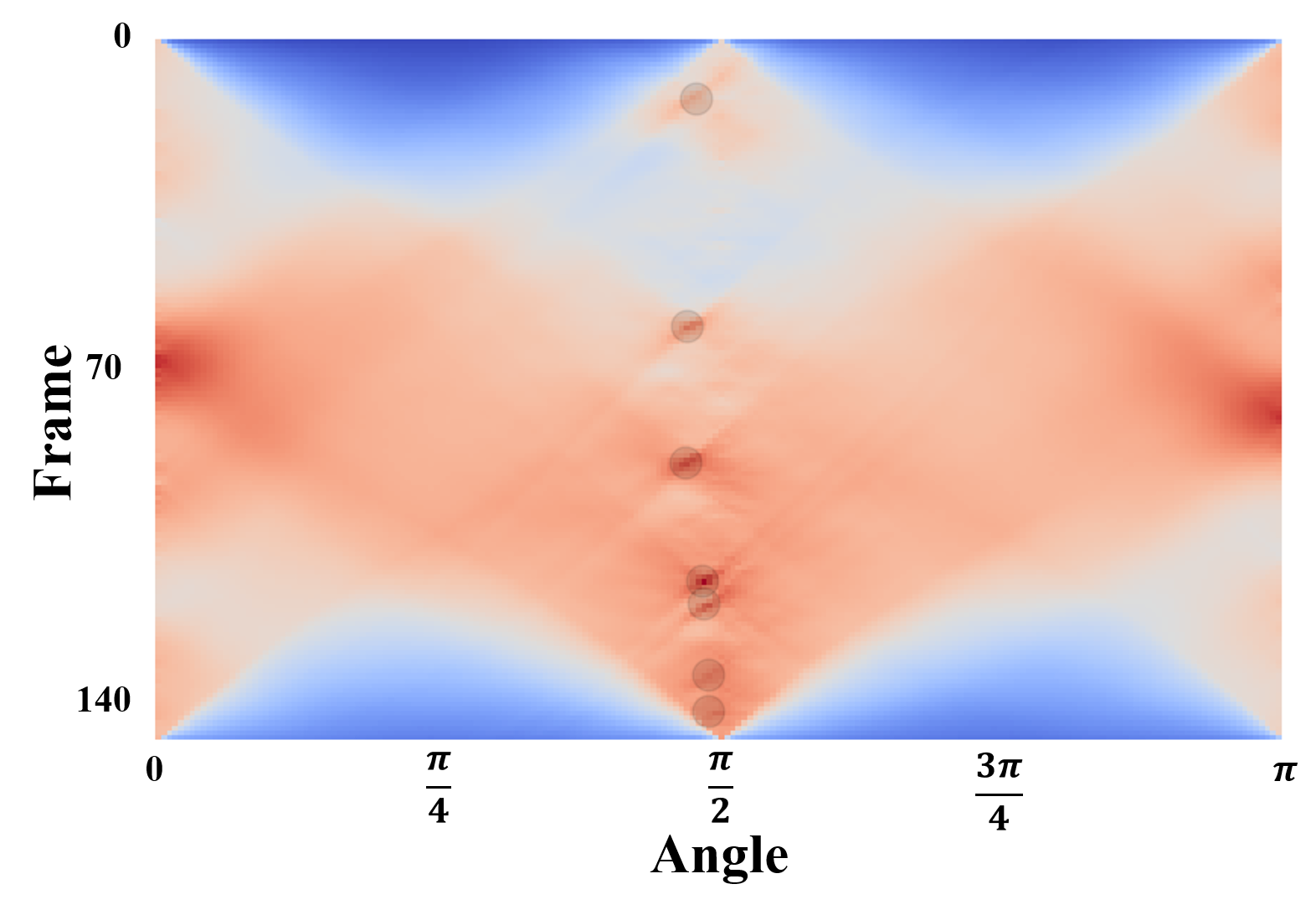}}

  \centerline{(b) Sinogram}\medskip
\end{minipage}
 % \vspace{-20px}
  \vspace{-10px}
\caption{\textbf{Detection of White Blood Cells (WBCs) in Our Nailfold Capillary Video Analysis System.} Visualization of WBC traversal through a given capillary as bright bands in (a), with corresponding detected local maxima annotated in a black circle in (b).}
\label{fig:velocity}
  \vspace{-15px}
\end{figure}

\vspace{-5px}
Utilizing nailfold videos from our dataset, our end-to-end video analysis pipeline aims to automatically measure the count of White Blood Cells (WBCs) traversing a given capillary, along with their speed as an approximate measure of blood flow velocity (see Figure \ref{fig:pipeline}). This process is further demonstrated through a specific case in Figure \ref{fig:velocity}.

%-------

Our method involves five primary steps. Initially, we achieve video stabilization using keypoint feature matching. Then morphological skeletonization is applied based on U-Net segmentation, and the average cross-sectional value forms the spatio-temporal profile of the specified capillary. Building upon a successful approach in prior work\cite{bourquard_analysis_2015}, we also apply a discrete-domain Radon transform to the spatio-temporal profile of the given capillary. This step enables the detection of WBC events by identifying local maxima on the sinogram, as the Radon transform maps 2D lines to peaks at specific positions, providing a convenient and noise-robust approach to WBC detection and analysis. In the final stage, we employ the Radon inversion transform to calculate the speed and occurrence time of the detected WBC events.

%--------
\begin{table}[bht]
    \centering
    \begin{threeparttable}
        \caption{Evaluation of Diameter Prediction at Subject Level}
        \begin{tabularx}{\columnwidth}{l*{3}{>{\centering\arraybackslash}X}}
            \toprule
            \textbf{Metric} & \textbf{MAE\tnote{1}} & \textbf{RMSE\tnote{1}} & \textbf{\#Subject\tnote{2}} \\
            \midrule
            Venous Limb & 0.989 & 1.230 & 54 \\
            Arterial Limb & 0.849 & 1.071 & 54 \\
            Apical Diameter & 1.674 & 2.023 & 52 \\
            \bottomrule
        \end{tabularx}
        \begin{tablenotes}\footnotesize
            \item[1] Measurements are reported in pixels.
            \item[2] Subjects with insufficient clear capillaries are marked with NaN values and excluded from the evaluation. 
        \end{tablenotes}
    \label{tab:regression_metrics}
\end{threeparttable}
\end{table}
%-------
\begin{table}[bht]
    \centering
    \begin{threeparttable}
        \caption{Evaluation of Abnormality Diagnosis for Each Measured Feature}
        \small
        \begin{tabularx}{\columnwidth}{lccccc}
            \toprule
            \textbf{Metric} & \textbf{Acc.} & \textbf{Prec.} & \textbf{Rec.} & \textbf{F1} & \textbf{\#Subject}\\
            \midrule
            Crossing Portion & 0.909 & 0.963 & 0.909 & 0.935 & 55 \\
            Tortuous Portion & 0.889 & 1.000 & 0.889 & 0.941 & 54\\
            Venous Limb & 0.759 & 0.708 & 0.759 & 0.719 & 54\\
            Arterial Limb & 0.667 & 0.630 & 0.667 & 0.623 &  54\\
            Apical Diameter & 0.654 & 0.604 & 0.654 & 0.603 & 52\\
            Length & 0.736 & 0.776 & 0.736 & 0.747 & 53\\
            \bottomrule
        \end{tabularx}
        \begin{tablenotes}\footnotesize
            \item[1] Normal ranges for each feature: Crossing Portion \(\leq 0.3\), Tortuous Portion \(\leq 0.1\), Venous Limb [11,17] \(\mu m\), Arterial Limb [9,13] \(\mu m\), Apical Diameter [12,18] \(\mu m\), Length of Capillary [150,250] \(\mu m\).
            \item[2] Acc. = Accuracy, Prec. = Precision, Rec. = Recall.
        \end{tablenotes}
        \label{tab:classification_metrics}
    \end{threeparttable}
\end{table}

%--------

\vspace{-10px}
\subsection{Experimental results}
\vspace{-5px}
To evaluate the performance of our fine-tuned deep-learning models on tasks such as segmentation, keypoint detection, and abnormality classification, we tested our models on separate test datasets using common standardized metrics, as detailed in Table \ref{tab:model_metrics}. For object detection, we prioritized sensitivity, while accuracy was the primary metric for classification tasks.

Furthermore, we conducted a comparative analysis with clinical benchmarks, where the average outcomes for each subject are derived from all corresponding images.
The accuracy of our model's diameter predictions at the subject level was evaluated through Mean Absolute Error (MAE) and Root Mean Square Error (RMSE), with results presented in Table \ref{tab:regression_metrics}. Lower values of MAE and RMSE indicate superior precision in measurements, underscoring the reliability of our approach.
Additionally, subject-level diagnosis of abnormalities was categorized as either ``normal'' or ``abnormal'', guided by clinical standards for each examined feature. This classification used key metrics such as accuracy, precision, recall, and F1 score to demonstrate the effectiveness of our automated nailfold capillary analysis system in a clinical context, as illustrated in Table \ref{tab:classification_metrics}.

In Table \ref{tab:regression_metrics}, our automated analysis algorithms demonstrated high precision, achieving mean absolute errors of $0.849$, $0.989$, and $1.674$ pixels for predicting arterial and venous limb widths and apical diameter, respectively.
For morphological abnormality assessment, our model achieved accuracies of $90.9\%$ and $88.9\%$ in predicting crossing and tortuous abnormalities, respectively, at the subject level, as demonstrated in Table \ref{tab:classification_metrics}. It's essential to note that clinical reports may contain subjective bias, with experts possibly inclined to classify cases as ``normal'', This inclination could account for any observed discrepancies between our results and clinical measurements.

\section{Discussion}
\vspace{-5px}

Our robust pipeline, enhanced by open-source codes and extensive datasets is not without limitations though. One key challenge is the realistic complexity introduced by our dataset’s images and videos, which closely mimic real-world scenarios. This complexity can potentially impact the effectiveness of existing deep-learning models, necessitating ongoing refinements. Additionally, our current data annotation process exhibits some inconsistencies with noise labels, indicating a need for a more adaptable method to address that issue.   

There are opportunities for improvement in our work, including integrating comprehensive algorithms and models, exploring a broader set of dynamic features such as microemboli, and refining both the overall pipeline efficiency and precision.  Our pipeline establishes a foundational benchmark, supported by an open-source dataset and codes, to facilitate and inspire future research in this domain. 

Besides, our research paves the way for future studies to explore the complex relationships between various capillary features and individual health conditions. The potential broader impact of our research, especially in the realm of personalized healthcare and mobile technology, is significant. Adapting our pipeline for use on mobile devices opens up exciting possibilities, allowing individuals to monitor their physical health conveniently, non-invasively, and affordably.

\section{Conclusion}
\vspace{-5px}
In conclusion, our research contributes to the field of medical imaging with the development of a state-of-the-art, fully automated nailfold capillary analysis pipeline. Key contributions of our study are outlined as follows:
(1) Comprehensive Dataset: We compiled an extensive dataset that includes 321 capillaroscopy images, 219 videos, and 68 clinical reports, all enriched with expert annotations for accurate segmentations and keypoints.
(2) Open-source Pipeline: To the best of our knowledge, we are the first to introduce and open-source a fully autonomous analytical framework that processes both images and videos, directly yielding precise metrics.
(3) Accurate Evaluation Results: Our system achieves high precision on average, with measurements of arterial and venous limb width accurate to less than 1 pixel, and an average $89.9\%$ accuracy in identifying key capillary patterns, as validated against clinical benchmarks.
Our study underscores the effectiveness of deep-learning models in advancing medical diagnostics and sets a solid groundwork for future research.

% -------------------------------------------------------------------------
\vfill
\pagebreak

\section{Compliance with Ethical Standards}
\vspace{-10px}
\label{sec:ethical}
This study was performed in line with the principles of the Declaration of Helsinki. Approval was granted by the Ethics Committee of my institution.

\vspace{-10px}
\section{Acknowledgments}
\vspace{-10px}
\label{sec:acknowledgments}
This work is supported by the Natural Science Foundation of China (NSFC) under Grant No. 62132010 and No. 62272055, Young Elite Scientists Sponsorship Program by CAST under Grant No 2021QNRC001, Beijing Natural Science Foundation under Grant No. QY23124, New Cornerstone Science Foundation through the XPLORER PRIZE, Beijing Key Lab of Networked Multimedia.
\vspace{-10px}
% References should be produced using the bibtex program from suitable
% BiBTeX files (here: strings, refs, manuals). The IEEEbib.bst bibliography
% style file from IEEE produces unsorted bibliography list.
% ------------------------------------------------------------------------- 
% \setlength{\bibsep}{0.5\baselineskip}
\begin{small} % Change the font size
    \bibliographystyle{IEEEbib}
    \bibliography{refs}
\end{small}

\end{document}